%Paper: hep-th/9209099
%From: jds@rademacher.math.upenn.edu
%Date: Thu, 24 Sep 92 14:39:00 GMT-0400

\input amstex
\input amssym.def
\magnification = \magstep1
\documentstyle{amsppt}

\NoBlackBoxes
\magnification=\magstep1
\def\bw{\bigwedge}
\def\16N{d_1\lbrack\phi_1\dots\phi_N\rbrack \pm \Sigma_1^N\pm\lbrack\phi_1
\dots
d_1 \phi_i\dots\phi_N\rbrack = \Sigma^{N-1}_{J=2}\pm\lbrack\lbrack\phi_{i_1}
\cdots \phi_{i_J} \rbrack\phi_{i_{J+1}}\cdots\phi_{i_N}\rbrack}
\def\cite#1{[{\bf #1}]}
 
\def\cite#1{[{\bf #1}]} 
%\pageno=1
% \headline={\ifnum\pageno>1 \tenbf
% SHLie algebra for physics \hss Lada-Stasheff\else\hfil\fi}
\line {\hss UNC-MATH-92/2}
 \line {\hss originally April 27, 1990, revised September 24, 1992}
\vskip3ex
\topmatter
\title  Introduction to sh Lie algebras for physicists \endtitle
\author     Tom Lada \\ Jim Stasheff \endauthor
\thanks Second author supported in part by
NSF grants DMS-8506637 DMS-8901975 DMS-92 and grateful to the
University of Pennsylvania for hospitality during the final
stages of this paper. \endthanks
\address Math Dept, NCSU, Raleigh NC 27695-8205 \endaddress
\email lada\@math.ncsu.edu \endemail
\address Math Dept, UNC, Chapel Hill NC 27599-3250 \endaddress
\email jds\@math.unc.edu \endemail

\endtopmatter
\document
Much of point particle physics can be described in terms of Lie algebras
and their representations.
Closed string field theory, on the other hand,
leads to a generalization of Lie algebra which arose naturally within
mathematics in the study of deformations of algebraic structures \cite{SS}.
It also appeared in work on higher spin particles \cite{BBvD}.
Representation theoretic analogs arose in the mathematical analysis of
the Batalin-Fradkin-Vilkovisky approach to constrained Hamiltonians \cite{S6}.
\par
The sh Lie algebra of closed string field theory
\cite {SZ}, \cite {KKS}, \cite K, \cite{Wies},  \cite{WZ}, \cite Z
is defined on the full Fock complex of the
theory, with the BRST differential $Q$.
Following Zwiebach \cite Z, we stipulate that the string fields
$B_1,B_2, \cdots$ are elements of
${\Cal H},$ the Hilbert space of the combined conformal
field theory of matter and ghosts.
The (n-fold) string product (bracket) for genus $0$ is denoted by
$$ \bigl[ B_1 , B_2, \cdots , B_n \bigr]_0 . $$
It has $n$ entries, that is, $n$ states in $\Cal H$.
Since we will deal only with genus $0$, we will omit the subscript $0$
henceforth.  The basic equation relating these brackets and
the BRST operator is:
$$
0= Q \bigl[ B_1 ,\cdots , B_n \bigr]
+\sum_{i=1}^n \pm \bigl[ B_1 , \cdots , QB_i , \cdots , B_n \bigr]
+ \sum \sigma (\  ,\  ) \bigl[ B_{i_1},\cdots, B_{i_j},
[B_{i_{j+1}}, \cdots , B_n]\,\bigr],  \tag1
$$
where the second sum is over all unshuffles (see below) and $\sigma(\ ,\ )$
denotes an appropriate sign.
\par

In their work on higher-spin particles, Berends, Burgers and van Dam consider
infinitesimal gauge transformations of the form
$$
\delta_{\xi}\phi = \partial\xi + \Sigma_{n=2}^{\infty}\ g^nT_n(\phi , \xi )
$$
where $T_n$ is $n$-linear in $\phi$ and linear in $\xi$ and $g$ is a
``coupling constant''.  Notice the unusual $\phi$-dependence of the
transformation.  They consider the case in which the commutator of
two such transformations is again of this form, which they use to define
$$
[\xi_1 , \xi_2 ] = \Sigma_{n=0}^{\infty}\ g^{n+1}C_n(\phi, \xi_1, \xi_2)
$$
where $C_n$ is $n$-linear in $\phi$ and linear in each $\xi_i.$
For this bracket to satisfy the Jacobi identity implies a whole sequence of
conditions for each order $g^n$.  In particular, if the $\xi_i$ and $\phi_j$
are fields of the same sort and we take $C_n = T_{n+1}$, then to order $g^2$
they find:
$$
\partial T_2(\phi_1,\phi_2, \phi_3) =
\sum_{i=1}^3 \pm T_2 ( \cdots , \partial\phi_i , \cdots )
+ \sum \pm T_1(T_1(\phi_{i_1}, \phi_{i_2}), \phi_{i_3})
$$
while to higher order, Burgers \cite B finds the full analogs of (1).
\par
To see the above formulas
as a generalization of those for a
differential graded Lie algebra is the major goal of this paper, hopefully
describing the mathematical essentials in terms
accessible to {\it physicists}.
\par
The concept of Lie algebra can be expressed in several different ways. The
most familiar are: in terms of generators and relations and in terms of a
bilinear ``bracket'' satisfying the Jacobi identity.
{\it In ``physical'' notation, let $X_a$ be a basis for V.  The
bracket $\lbrack \quad, \quad\rbrack$ can be specified by structure constants
$C^c_{ab}$ via the formula:
$$\lbrack X_a,X_b\rbrack = C^c_{ab} X_c.$$
 The structure constants are skew-symmetric in the lower indices $a,b$.}
\par
A much more subtle
description appears in the homological study of Lie algebras, but it is this
description which is the most useful in homological perturbation theory and in
mathematical physics in the guise of BRST operators for open algebras and in
the
algebra for closed string field theory.
This description is implicit in the somewhat  more familiar dual
formulation of the Chevalley-Eilenberg
cochain complex for Lie algebra cohomology \cite{CE}: An {\bf n-cochain} is a
skew-symmetric n-linear function $\omega: V\times \dots\times V\to \Bbb R$ and
the coboundary $d\omega$ is defined by
$$
d\omega (v_1,\dots ,v_{n+1}) =
\underset{i<j}\to\Sigma (-1)^{i+j} \omega ([v_i, v_j], v_1,\dots ,
v_1,\dots \hat v_i \dots \hat v_j \dots , v_{n+1})
$$
where the ``hatted'' variables are to be omitted. {\it With respect to a
dual basis $\omega^a$ of the dual vector space $V^*$, we can write
$d$ as $- 1/2\ \omega^a\omega^b C^c_{ab}\partial_{X_{c}}.$}
\par
  We can deal directly with
the vectors rather than with the multi-linear `forms' at the expense
of introducing a new point of view and consideration of skew-symmetric
tensors ({\it multi-vectors}).
\par
\vskip2ex\noindent
A Lie algebra is equivalent to the following data:
\vskip0.5ex
A vector space V (assumed finite dimensional for simplicity  of exposition).
\vskip0.5ex
The skew (or alternating) tensor products of V, denoted
$$\bw     V = \{\bw ^nV\}.$$
A linear map
$$d:\bw     V \rightarrow \bw     V$$
 which lowers $n$ by one and is a {\it co-}derivation such that $ d^2 = 0$.
\vskip0.5ex\noindent
(That $d$ is a {\bf co-derivation} means just that
$$d(v_1\wedge \dots \wedge v_n) = \underset{i<j}\to\Sigma (-1)^{i+j}
d(v_i\wedge v_j)\wedge v_1\wedge \dots \hat v_i \dots \hat v_j \dots
\wedge v_n.$$
{\it For example, $$d(X_a\wedge X_b\wedge X_c) = -C^e_{ab}X_e\wedge X_c +
C^e_{ac}X_e\wedge X_b -C^e_{bc}X_e\wedge X_a.\ )$$}

\par
It may not be immediately obvious, but $d$ restricted to $\bw ^2$ is
to be interpreted
as a bracket: $d(v_1\wedge v_2) = [v_1, v_2]$ and $d^2 = 0$ is equivalent to
the
Jacobi identity.
\par
{}From the point of view of the skew tensor powers of $V$,
an {\bf sh Lie algebra (strongly homotopy Lie algebra)}
is similarly equivalent to a  straightforward generalization in which d is
 replaced by  a co-derivation
$$D = d_1 + d_2 +d_3 + \dots$$ where $d_i$ lowers $n$ by $i-1,$ in particular,
$d_n(v_1\wedge \dots \wedge v_n)\in V$.
We say that $D$ is a coderivation to summarize the several conditions:
$$
d_i(v_1\wedge \dots \wedge v_n) = \Sigma \pm d_i(v_{i_1}\wedge\dots\wedge
v_{i _j})\wedge v_{i_{j+1}}\wedge \dots\wedge v_{i_n}
$$
where the sum is over
all {\it unshuffles} of $\lbrace 1,\dots ,n\rbrace$,
that is, all permutations that keep $i_1,\dots ,i_j$ and $i_{j+1},\dots ,i_n$
in the same relative order.  A {\it shuffle} of two ordered sets (decks
of cards) is a permutation of the ordered union which preseves the order
of each of the given subsets; an {\it unshuffle} reverses the process, cf.
Maxwell's demon. (The serious issue of signs will be addressed
below.)
\vskip2ex
Notice that the old $d$ corresponds to $d_2$ since
$d(v_1\wedge v_2) = [v_1, v_2]\in V.$
On the other hand, for the new D, the component $d_2$
no longer is of square zero by itself and hence corresponds to a bracket which
does  NOT necessarily satisfy the Jacobi identity.  Let us look in detail at
what can happen instead:
\vskip1ex
Expand $D^2 = 0$ in its homogeneous components and set them separately
equal to zero.  We have then:
$$
0)\quad  d_1^2 = 0
$$
so $(V,d_1)$ is a complex or differential (graded) module.  (Usually $V$
is itself a graded module and $D$ is a {\it graded} co-derivation, which
implies appropriate signs in applying $D$ to $x_1\wedge \dots \wedge x_n$.
Also $d_1$ typically raises (or lowers) that `internal'
degree by 1.)  {\it Perhaps the most important examples in physics are the
exterior derivative on differential forms and the BRST operator on suitable
fields, whether physical or ghostly.})
$$
1) \quad d_1 d_2 + d_2 d_1 = 0
$$
so, with appropriate sign conventions,  $d_2$
gives a bracket $[v_1, v_2]\in V$ for which
$d_1$ is a derivation.
$$
2) \quad d_1 d_3 + d_2 d_2 + d_3 d_1 = 0
$$
or equivalently
$$
d_2 d_2 = -(d_1 d_3 + d_3 d_1).
$$
If we further adopt the notation:
$$
d_3(x_1\wedge x_2\wedge x_3) = \lbrack x_1,x_2,x_3\rbrack,
$$
then we have
$$
[[v_1, v_2], v_3] \pm [[v_1, v_3], v_2] \pm [[v_2, v_3], v_2] =
- d_1[v_1, v_2, v_3] \pm [d_1v_1, v_2, v_3]\pm[v_1, d_1v_2, v_3]\pm [v_1, v_2,
d_1v_3].
$$
{}From this, with use of the skew-symmetry of $[\ ,\ ]$,
 we see that now the Jacobi identity holds modulo
the right hand side.  {\it In physical language, the Jacobi identity holds
modulo a BRST exact term.} In the language of homological algebra,
$d_3$ is a chain homotopy,  so we say that $(V,d_2)$ satisfies the Jacobi
identity {\it up to homotopy} or $(V,d_1,d_2,d_3)$ is a
{\bf homotopy Lie algebra}.  The adverb ``{\bf strongly}'' is added to refer
to the other $d_i$.
\vskip2ex
{\it In physical notation, restricting $d_3$  to $\bw^3 V$, we can write:
$$
d_3(X_a\wedge X_b\wedge X_c) = C^e_{abc} X_e
$$
where $C^e_{abc}$ is skew-symmetric
in the lower indices. Similarly, we can write
$$
d_1X_a = C_a^b X_b.
$$
Just as the Jacobi identity can be written as a quadratic
equation in the $C^c_{ab},$ so equation 2) can be written as a quadratic
equation in the $C_a^b,\, C^c_{ab}$
and $C^e_{abc}$.}
\par
If we adopt the notation that $d_1 = Q$ and in general:
$$
d_n(x_1\wedge \dots\wedge x_n) = \lbrack x_1,x_2,\dots ,x_n\rbrack,
$$
then the appropriate homogeneous piece of $D^2 = 0$ is (up to sign conventions
and up to some constants
related to conventions on the definition of $\bw V$) precisely the equation
(1) occurring in the `non-polynomial' version of
the genus zero closed string field algebra:
The sum is not only over j but also over
all $(j,n-j)$ unshuffles.
Making the  correspondence precise, including the approriate signs,
requires some care, as in the next section.
\par
In the higher spin particle algebra of \cite{BBvD}, variations $\delta_
{\epsilon}$ do not respect a strict bracket $[\epsilon_1, \epsilon_2]$
but rather an sh Lie structure on the space of $\epsilon$'s.  In the
Batalin-Fradkin-Vilkovisky operator for constraints forming an `open' algebra
with structure functions, one sees a similar structure \cite {S6}.
% (see below).
\par
This paper is organized as follows:  After establishing our notation and
conventions (especially for signs), we give the formal definition
of sh Lie structure and verify the equivalence with
a formulation in terms of a `nilpotent' operator on $\bw sV.$
Comparison with the physics literature calls attention to some further
subtleties of signs. Then we
establish the sh analog of the familiar fact that commutators in an
associative algebra form a Lie algebra.  We next point out the relevance
of these
structures to $N+1$-point functions in physics.  We remark on the
distinction between these structures on the cohomology level and at the
underlying form level  and conclude with the basic theorem of
Homological Perturbation Theory relating higher order bracket operations on
cohomology to strict
Lie algebra structures on forms.
\par
Recognition that the mathematical structure of  sh Lie algebras was
appearing in physics first occurred in discussions with Burgers and
van Dam and then
with Zwiebach at the last GUT Workshop.  He called our attention
to the relevant preprints \cite {SZ} and \cite {KKS} and has been
a continued source of inspiration for the present paper, cf. especially
\cite Z.   Our attempts to make this exposition accessible to physicists
have benefitted also from comments of Henrik Aratyn, Jos\'e Figueroa-O'Farrill
and Takashi Kimura.
We are indebted to Elizabeth Jones Dempster for some of the hard
technical details with regard to signs, especially in comparing sh Lie
structures to sh associative ones.
%\head sh Lie Structures\endhead
\subhead Notation and Conventions \endsubhead\par
In dealing with maps of differential graded vector spaces, it is crucial to
keep careful track of signs.  First, there are the appropriate signs for
any graded (or {\it super}) context. The basic convention is that whenever
two symbols,
of degree m and n respectively, are interchanged, a sign of $(-1)^{mn}$ is
introduced.  In particular, if $\sigma$ is a permutation that is acting on a
string of such symbols, $e(\sigma)$ will denote the sign that results from
iteration of the basic convention, i.e. $e(\sigma) = (-1)^k$ where $k$
is the number of interchanges of odd symbols.  We also let $(-1)^\sigma$
denote the usual
sign of the permutation $\sigma$. It is important to note that
$e(\sigma)$ does not involve $(-1)^\sigma$
as a factor.  However, if all the elements in question have degree 1, then
$e(\sigma) = (-1)^\sigma$.
\par
We always regard the symmetric group $\Cal S_n$ as acting on $V^{\otimes n}$ by
$$\sigma (v_1\otimes\dots\otimes v_n) = v_{\sigma (1)}\otimes \dots\otimes
 v_{\sigma (n)}.$$
A map of graded vector spaces $f:V^{\otimes n}\to W$is called
$$\text{{\bf symmetric} if\ \ } f(v_{\sigma (1)}\otimes \dots\otimes
v_{\sigma (n)})= e(\sigma)f(v_1\otimes  \dots\otimes v_n)$$
and $$\text{{\bf skew symmetric} if\ \ } f(v_{\sigma (1)}\otimes \dots\otimes
  v_{\sigma (n)})=
 (-1)^{\sigma}e(\sigma)f(v_1,\dots ,v_n)$$ for all $\sigma \in \Cal S_n.$
{\it In terms of a basis for $V$, we can write $f$ in terms of its
coefficients $f^{a_1\dots a_n}$, then symmetric or skew symmetric has the
usual interpretation with respect to permutations of $a_1\dots a_n$.}
\par
Now let $V=\{V^i\}$ be a graded vector space over  a field $k$.
Let $T^*(V)$
denote the tensor vector space generated by V,
i.e. $\{ V^{\otimes n}\} $.  We do NOT consider
$T^*(V)$
as the tensor algebra, but rather as a {\bf coalgebra} with
 the standard coalgebra structure given by the diagonal $$
 \bigtriangleup (v_1\otimes \dots \otimes v_n)=
\sum_{j=0}^n (v_1\otimes \dots \otimes v_j)\otimes (v_{j+1} \otimes\dots
\otimes v_n).
$$
(Here $V^{\otimes 0}$ is to be identified with $k$ and the terms with
$j = 0 $ or $j = n$ are of the form $1 \otimes (\dots )$ and $(\dots )\otimes
1$
respectively.)  The use of coalgebras is an efficient device for some
of our expressions, but it is sufficient to follow the argument in terms
of ordinary tensors.
\par
In particular, we will make use of $\bw V$
{\it which consists of the skew-symmetric tensors (in the graded sense)},
that is the subspace (in fact, sub-coalgebra) of
$T^*(V)$
which is fixed under the above action of the symmetric group on
$V^{\otimes n}$.
Although we will not need the terminology,
$\bw V$ is known as the {\bf free cocommutative coalgebra}
generated by V with reduced diagonal given by $$
\bar \bigtriangleup (u_1 \wedge\dots\wedge u_n)=\sum_{j=1}^{n-1}
\sum_\sigma e(\sigma)(u_{\sigma (1)}\wedge\dots\wedge u_{\sigma (j)})\otimes
(u_{\sigma (j+1)} \wedge\dots\wedge u_{\sigma (n)})$$ where $\sigma$
runs through all $(j,n-j)$ unshuffles.
\subhead sh Lie Structures\endsubhead\par
Let V be a differential graded vector space with differential denoted by
$l_1: V_i\to V_{i-1}.$   We first recall the graded version
of an ordinary Lie algebra.
\definition{Definition} A {\bf graded Lie algebra} is a graded vector
space $V = \{V_p\}$ together with a graded skew commutative bracket
$[\ ,\ ]: V\otimes V\rightarrow V$ such that
$V_p\otimes V_q\to V_{p+q}$
satisfying the graded Jacobi identity:
$$
[u, [v,w]] = [[u,v],w] + (-1)^{pq}[v,[u,w]].
$$
\enddefinition
An alternative ``with the grading reduced by one'' (meaning
$V_p\otimes V_q\to V_{p+q-1}$) occurs in the Hochschild cohomology
of an associative algebra as part of the structure of a Gerstenhaber
algebra \cite G, a structure which is also making an appearance in physics.
\par
Examples have existed for a long time.  Perhaps
the earliest is the Schouten bracket of multivectors although it was
not identified as such until long after its introduction.  Similarly the
Whitehead product of homotopy groups \cite {Wh} (which has the reduced grading)
and the corresponding
Samelson product \cite {Sa} were also not so  identified initially.
\par
We will also consider
the differential graded vector space $sV$ which is defined by
$(sV)_i=V_{i-1}$ with $\hat l_1(sv) = - sl_1(v)$.  The use of such a
suspension operator $s$ is implicit in the Chevalley-Eilenberg complex,
but for the more complicated sh Lie structures, it
is best to make it explicit. For similar reasons, especially subtlety of
signs, and also for
comparison with the corresponding sh associative algebras, we have chosen to
express this
section in terms of maps $l_n$ and $\hat l_n$ rather than in terms of
the $d_n$ of the introduction.
\definition{Definition} An {\bf sh Lie structure} on V is a collection of skew
symmetric linear maps $l_n:\bigotimes ^nV \longrightarrow V$ of degree $2-n$
(for cochain complexes, and $n-2$ for chain complexes)
such that
 $$\sum_{i+j=n+1} \sum_{\sigma} e(\sigma) (-1)^\sigma \alpha(i,j)\,l_i (l_j
(v_{\sigma (1)}\otimes \dots \otimes v_{\sigma (j)}) \otimes v_{\sigma(j+1)}
\otimes \dots\otimes v_{\sigma (n)})=0\tag2$$
 where $\alpha (i,j)=-1$ only if $i$ is odd and $j$ is even and $1$ otherwise,
while $\sigma$
runs through all $(j,n-j)$ unshuffles (cf. \cite J).
\enddefinition
\par It is clear that we may use the skew symmetry of the $l_n$'s to induce
linear maps $l_n :\bigwedge^nV\longrightarrow V$ that satisfy the
same equations
in the definition.  We note that the map $l_2$ may be viewed as a usual
(graded)
Lie bracket and that when $n=3$ and $l_3=0$, the definition yields the usual
(graded) Jacobi identity.  In general, $l_3$ is a homotopy between the Jacobi
expression and $0$ while
the other $l_n$'s are higher homotopies.
\par
It is worth calling attention to the matter of degrees, both for the forms
(e.g. {\it ghost degree in the BRST context}) and the degree
of the $N$-fold operations.  The original mathematical
formulation as above) of sh Lie algebras (and their sh associative
algebra predecessors) assumed that the (2-fold) product or bracket
had degree zero, i.e. the degree of the bracket
was the sum of the degrees of the factors and that
the degree of $d = d_1$ was 1 (for cohomology) or -1
(for homology).  The defining equality (1) then
determines the degrees of the other $N$-fold brackets.
If the bracket should itself have a degree (for example, -1 for the
Whitehead product or the Gerstenhaber bracket or 1 in some physical examples),
then the degrees of the other $N$-fold operations would be adjusted
accordingly, as would the signs in (2).  In particular, as Zwiebach is careful
to point out in section 4.1 of \cite Z, in closed string field
theory, the degree is given by the statistics, so the degree of the
2-fold bracket is 1.  As he explains in detail, a simple shift in
counting degrees and appropriate changes in signs establishes
the equivalence of our two sets of conventions.  This is precisely the
shift that relates the Whitehead product to the Samelson product
\cite{Wh}, \cite {Sa}.
\par
We now use the graded vector space $sV$; recall $(sV)_i=V_{i-1}$.
Define linear maps $\hat l_n:\bigwedge ^nsV \longrightarrow sV$
 by $$
\hat l_n(sv_1\wedge\dots\wedge sv_n)=\left\{\aligned
(-1)^{\sum_{i=1}^{n/2}\vert
 v_{2i-1}\vert}\,sl_n(v_1\wedge\dots\wedge v_n), \qquad n
\quad\text{even}\\ -(-1)^{\sum_{i=1}^{(n-1)/2}\vert
v_{2i}\vert}\,sl_n(v_1\wedge
\dots\wedge v_n), \qquad n \quad\text{odd}\endaligned\right\}\tag3$$where
$\vert
 v_k\vert = $ degree of $v_k$.
\par
We may now further extend $\hat l_n:\bigwedge ^ksV
\longrightarrow\bigwedge^{k-n+1}sV$ as a coderivation with respect to the usual
coproduct on $\bigwedge^*sV$.  Explicitly, we have
$$\hat\l_n(sv_1\wedge\dots\wedge sv_k)=
\sum_\sigma e(\sigma)\,\hat l_n(sv_{\sigma(1)}\wedge\dots\wedge
sv_{\sigma(n)})\wedge sv_{\sigma(n+1)}\wedge\dots\wedge sv_{\sigma(k)}\tag4$$
where $\sigma$ runs through all $(n,k-n)$ unshuffles. Define the degree of
an element in $\bigwedge ^*sV$ by $\vert sv_1\wedge \dots \wedge sv_k \vert
= k+\sum_{i=1}^k \vert v_i \vert$.  Note that with this grading each
$\hat l_n$ has degree $-1$.  The ordering convention in (4) with the
$\hat l_n$ terms listed first provides a convenient way of keeping track
of the signs and the unshuffles.  We observe however that if an element of
$\bigwedge ^ksV$ has the form $$sv_{\sigma(1)}\wedge \dots \wedge
sv_{\sigma(i)}\wedge sv_{\sigma(i+1)}\wedge \dots \wedge sv_{\sigma(i+n)}
\wedge \dots \wedge sv_{\sigma(k)}$$ where the sequence $\sigma(i+1),
\dots \sigma(i+n) $ respects the original ordering of the $sv_i$'s,
we then have among the summands in (4)
 an equivalent term of the form
$$(-1)^{(\sum_{p=1}^i\vert sv_{\sigma(p)}\vert)(\vert \hat l_n\vert )}
e(\sigma) sv_{\sigma(1)} \wedge \dots \wedge sv_{\sigma(i)}\wedge
\hat l_n(sv_{\sigma(i+1)} \wedge \dots \wedge sv_{\sigma(i+n)}) \wedge
\dots \wedge sv_{\sigma(k)}$$ by use of the basic sign conventions.
\par
In order to calculate compositions of the $l_n$'s we require the
following lemma.
\proclaim{Lemma} $\hat l_i(\hat l_j(sv_1\wedge\dots\wedge sv_j)\wedge sv_{j+1}
\wedge\dots\wedge sv_{j+i-1})=$
$$(-1)^{\sum_{p=1}^{i+j-2/2}\vert v_{2p}\vert}sl_i(l_j(v_1\wedge\dots
\wedge v_j)\wedge v_{j+1}\wedge\dots\wedge v_{i+j-1})\text{ if }i+j
\text{ is even}$$
$$(-1)^{\sum_{p=1}^{i+j-1/2}\vert v_{2p-1}\vert}sl_i(l_j(v_1\wedge\dots
\wedge v_j)\wedge v_{j+1}\wedge\dots\wedge v_{i+j-1})\text{ if }i
\text{ is even and }j\text{ is  odd}$$
$$(-1)^{1+\sum_{p=1}^{i+j-1/2}\vert v_{2p-1}\vert}sl_i(l_j(v_1\wedge
\dots\wedge v_j)\wedge v_{j+1}\wedge\dots\wedge v_{i+j-1})\text{ if }
j\text{ is even and }i\text{ is  odd.}$$
\endproclaim
\demo{Proof} We first assume that both
$i+j$ and $j$ are even.  Then by using (3), we have
$$\hat l_i(\hat l_j(sv_1\wedge\dots\wedge sv_j)\wedge sv_{j+1}\wedge\dots
\wedge sv_{i+j-1})=$$
$$(-1)^{\sum_{p=1}^{j/2}\vert v_{2p-1}\vert}(-1)^{\vert l_j(v_1\wedge\dots
\wedge v_j)\vert + \sum_{p=1}^{i-2/2}\vert v_{j+2p}\vert}sl_i(l_j(v_1\wedge
\dots\wedge v_j)\wedge v_{j+1}\wedge\dots\wedge v_{i+j-1}).$$
Note that the sign in the above expression is equal to
$$(-1)^{\sum_{p=1}^{j/2}\vert v_{2p-1}\vert}(-1)^{j-2+\sum_{p=1}^j\vert
v_p\vert +\sum_{p=1}^{i-2/2}\vert v_{j+2p}\vert}=(-1)^{\sum_{p=1}
^{i+j-2}\vert v_{2p}\vert}$$
which is what is claimed.  The other three cases follow from a
similar calculation.\qed\enddemo
\proclaim{Theorem} Let $D:\bigwedge^*sV\longrightarrow\bigwedge^*sV$ be given
by
$D=\sum_i\hat l_i$.  Then $D^2=0.$
\endproclaim
{\bf Note:}  This gives a generalization of the Chevalley-Eilenberg complex
 for
a Lie algebra.  The rest of this section is devoted to a proof which amounts
to checking that the signs cancel appropriately.
\demo{Proof} Evaluate $D^2= \sum_{k=2}^{n+1} \sum_{i+j=k} \hat l_i \hat
l_j$ on $sv_1 \wedge \dots \wedge sv_n$.  In fact we claim that for each
$k\leq n+1$, we have $\sum_{i+j=k}\hat l_i \hat l_j(sv_1 \wedge \dots
\wedge sv_n) =0$.  We use (4) to write
$$
\hat l_i \hat l_j (sv_1 \wedge \dots \wedge sv_n)
=\hat l_i(\sum_{\sigma}e(\sigma) \hat l_j(sv_{\sigma(1)}\wedge \dots \wedge
sv_{\sigma(j)}) \wedge sv_{\sigma(j+1)}\wedge \dots \wedge sv_{\sigma(n)}).
$$
Now use (4) again to evaluate $\hat l_i$.  At this point we note that a typical
summand of the resulting expression will have the form
$$e(\sigma) \hat l_i( \hat l_j(sv_{\sigma(1)} \wedge \dots \wedge sv_{\sigma
(j)})\wedge sv_{\sigma(j+1)} \wedge \dots \wedge sv_{\sigma(j+i-1)}) \wedge
sv_{\sigma(j+i)}\wedge \dots \wedge sv_{\sigma(n)} \tag a$$
or
$$e(\sigma) \hat l_i(sv_{\sigma(1)}\wedge \dots \wedge sv_{\sigma(i)})\wedge
 \hat l_j(sv_{\sigma(i+1)}\wedge \dots \wedge sv_{\sigma(i+j)})\wedge
sv_{\sigma(i+j+1)}\wedge \dots \wedge sv_{\sigma(n)} \tag b$$
\par\noindent
Here $\sigma$ refers to a composition of unshuffles and $e(\sigma)$ is the
resulting product of their signs.
\par
We begin with the type (a) terms.  First collect all terms together that
have identical last $n-j-i+1$ entries and denote this sequence of terms by
$sv_Q$.  Then
$$\sum_{i+j=k}\sum_{\sigma}e(\sigma)\hat l_i\hat l_j(sv_{\sigma(1)}\wedge
\dots\wedge sv_{\sigma(j)})\wedge sv_{\sigma(j+1)}\wedge\dots\wedge
sv_{\sigma(j+i-1)}\wedge sv_Q=$$
$$(-1)^{\sum_{p=1}^{i+j-2/2}\vert v_{2p}\vert}s(\sum_{i+j=k}\sum_{\sigma}
e(\sigma)(-1)^{\sigma}l_i(l_j(v_{\sigma(1)}\wedge\dots\wedge v_{\sigma(j)})
\wedge v_{\sigma(j+1)}\wedge\dots\wedge v_{\sigma(j+i-1)})\wedge sv_Q)
\tag 5$$
if $i+j$ is even, and
$$(-1)^{\sum_{p=1}^{i+j-2/2}\vert v_{2p-1}\vert}s(\sum_{i+j=k}\sum_{\sigma}
e(\sigma)(-1)^{\sigma}(-1)^il_i(l_j(v_{\sigma(1)}\wedge\dots\wedge v_{\sigma
(j)})\wedge v_{\sigma(j+1)}\wedge\dots\wedge v_{\sigma(j+i-1)})\wedge sv_Q)
\tag 6$$
if $i+j$ is odd.  Here $\sigma$ is taken over all $(j,i-1)$ unshuffles
in the first $j+i-1$ coordinates.
Both (5) and (6) are equal to $0$ by (2).
\par
To verify this claim we use
 the fact that each unshuffle is the product of transpositions, and
examine the effect of one transposition on the term
$$\hat l_i(\hat l_j(sv_1\wedge\dots\wedge sv_j)\wedge sv_{j+1}\wedge
\dots\wedge sv_{i+j-1})$$
while using the lemma  to evaluate $\hat l_i\hat l_j$.
In fact, the transposition
that we need consider is $\sigma = (j,j+1)$.  Consequently, we have
$$(-1)^{\vert sv_j\vert\vert sv_{j+1}\vert}\hat l_i(\hat l_j(sv_1\wedge\dots
\wedge sv_{j+1})\wedge sv_j\wedge\dots\wedge sv_{i+j-1})=$$
$$(-1)^{\vert v_j\vert\vert v_{j+1}\vert+1+\vert v_j\vert +\vert v_{j+1}\vert}
\hat l_i(\hat l_j(sv_1\wedge\dots\wedge sv_{j+1})\wedge sv_j\wedge\dots
\wedge sv_{i+j-1})=$$
$$e(\sigma)(-1)^{\sigma}(-1)^{\vert v_j\vert\vert v_{j+1}\vert}(-1)^{\sum
_{p=1}^{j+i-2/2}\vert v_{\sigma(2p)}\vert}sl_i(l_j(v_1\wedge\dots\wedge
v_{\sigma(j)})\wedge v_{\sigma(j+1)}\wedge\dots\wedge v_{i+j-1})$$
by the lemma when $i+j$ is even.  We note that since $\sigma(j)=j+1$ and
since $j$ and $j+1$ have opposite parity, the exponent
$$\vert v_j\vert +\vert v_{j+1}\vert +\sum_{p=1}^{i+j-2/2}
\vert v_{\sigma(2p)}\vert=
\sum_{p=1}^{i+j-2/2}\vert v_{2p}\vert.$$  The case in which $i+j$ is
odd follows by a similar calculation.
\par
We now turn our attention to the terms of type (b).  These terms occur
in pairs with opposite signs and thus cancel each other out.  They will
occur when the sequences $\sigma(1),\dots,\sigma(i)$ and $\sigma(i+1),
\dots,\sigma(i+j)$ are ordered with respect to the usual ordering on
the integers.  We then evaluate
$$\hat l_i\hat l_j(sv_{\sigma(1)}\wedge\dots\wedge sv_{\sigma(i)}\wedge
 sv_{\sigma(i+1)}\wedge\dots\wedge sv_{\sigma(i+j)}\wedge sv_Q)=$$
$$(-1)^{(\sum_{p=1}^{i}\vert sv_{\sigma(p)}\vert )\vert \hat l_j\vert}
\hat l_i(sv_{\sigma(1)}\wedge\dots\wedge sv_{\sigma(i)}\wedge\hat l_j(
sv_{\sigma(i+1)}\wedge\dots\wedge sv_{\sigma(i+j)})\wedge sv_Q=$$
$$(-1)^{\sum_{p=1}^i\vert sv_{\sigma(p)}\vert}\hat l_i(sv_{\sigma(1)}
\wedge\dots\wedge sv_{\sigma(i)})\wedge \hat l_j(sv_{\sigma(i+1)}\wedge
\dots\wedge sv_{\sigma(i+j)})\wedge sv_Q.$$
On the other hand,
$$\hat l_j\hat l_i(sv_{\sigma(1)}\wedge\dots\wedge sv_{\sigma(i+j)}
\wedge sv_Q)=$$
$$\hat l_j(\hat l_i(sv_{\sigma(1)}\wedge\dots\wedge sv_{\sigma(i)})\wedge
sv_{\sigma(i+1)}\wedge\dots\wedge sv_{\sigma(i+j)}\wedge sv_Q)=$$
$$(-1)^{(\sum_{p=1}^i\vert sv_{\sigma(p)}\vert)\vert \hat l_j\vert}
(-1)^{\vert\hat l_i\vert\vert \hat l_j\vert}\hat l_i(sv_{\sigma(1)}\wedge
\dots\wedge sv_{\sigma(i)})\wedge \hat l_j(sv_{\sigma(i+1)}\wedge\dots
\wedge sv_{\sigma(i+j)})\wedge sv_Q.$$
Since $\vert\hat l_k\vert =-1$, the signs are as claimed.  Note that
in the evaluation of $\hat l_i$ and $\hat l_j$ here, we were only
interested in the summands listed.\qed
\enddemo

\subhead Commutators in relation to the sh associative case\endsubhead
\par
Physicists often refer to Lie brackets as commutators because commutators
of elements in an associative algebra define a bracket satisfying the
axioms for a Lie algebra: $[x,y] = xy - yx$ or, in the graded case,
$[x,y] = xy - (-1)^{|x||y|}yx.$  There is a notion of strongly homotopy
associative (sha) algebra, much older than that of sh Lie algebra (cf. \cite
{S3,5}); it is natural to try to put
into the homotopy context
the canonical construction of a Lie algebra from an associative algebra
by use of commutators.
\definition{Definition} An {\bf strongly homotopy associative (sha) structure}
on V is a collection of
linear maps $m_n:\bigotimes ^nV \longrightarrow V$ of degree $2-n$
(for cochain complexes, and $n-2$ for chain complexes)
such that
$$
\sum_{i+j=n+1} \sum_k \beta(i,j,k)\,m_i(v_1\otimes \dots \otimes
v_{k-1} \otimes
m_j(v_k\otimes \dots\otimes v_{k+j-1}) \otimes\dots\otimes v_n)=0\tag7
$$

where $\beta(i,j,k)$ is the sign given by the parity of
$(j+1)k+j(n+\sum_{m=1}^{k-1}\vert x_m\vert).$
\enddefinition
\definition{Definition} Given a
differential graded vector space $V$ with an sha structure
$\{ m_n\}$, the {\bf commutator sh Lie structure} is defined on $V$ by
$$l_n(v_1\otimes\dots\otimes v_n)=\sum_{\sigma}(-1)^{\sigma}e(\sigma)
m_n(v_{\sigma (1)}\otimes\dots \otimes v_{\sigma(n)})$$
where the sum is taken over all permutations $\sigma\in S_n$.
\enddefinition
When
$n=2$, $l_2(v_1\otimes v_2)=m_2(v_1\otimes v_2)-(-1)^{\vert v_1\vert \vert
v_2\vert}m_2(v_2\otimes v_1)$ is the graded commutator.  The possible lack
of associativity of $m_2$ will in general prevent the Jacobi expression
$$l_2(l_2(v_1\otimes v_2)\otimes v_3)-(-1)^{\vert v_2\vert \vert v_3\vert}
l_2(l_2(v_1\otimes v_3)\otimes v_2)+(-1)^{\vert v_1\vert (\vert v_2\vert
+\vert v_3\vert)}l_2(l_2(v_2\otimes v_3)\otimes v_1)\tag8$$
from being equal to zero.  However, with $l_3(v_1\otimes v_2\otimes v_3)
=\sum_{\sigma \in S_3} (-1)^{\sigma}e(\sigma)
m_3(v_{\sigma(1)}\otimes v_{\sigma(2)}\otimes
v_{\sigma(3)})$, one may check that (8) is equal to $l_1l_3+l_3l_1$ where
$l_1$ is defined to be $m_1$.  These details, as well as those required
for $4$-tensors are made explicit by Elizabeth Jones Dempster in [J].
The general verification that the above defined `commutator' $l_n$'s satisfy
the defining equation for an sh Lie structure will appear elsewhere.
\subhead N+1-point functions \endsubhead\par
As a generalization of Lie algebras, sh Lie algebras
appear in closed string field theory as symmetries or gauge transformations.
The corresponding Lagrangians consist of (sums of)
$N+1$-point functions for all $N$ (cf.
\cite {SZ}, \cite {KKS}, \cite K, \cite{Wies}, \cite Z, \cite{WZ}).
They can be regarded as being formed from the $N$-fold brackets
$ \lbrack x_1,x_2,\dots ,x_N\rbrack, $
by evaluation with a dual field via an inner product or the $N+1$-point
functions can be described directly.  The latter is more appropriate for
Kontsevich's new invariants \cite{Kon}, but we will leave it to him to present
those details.  We adopt Dirac's
bra-ket notation and so write $\vert x>$ instead of just $x$.  We
then write $<x\vert$ for elements of the dual space.  In terms of
a basis $x_{\alpha}$ or rather $\vert x_{\alpha}>$, we have a dual
basis $<x_{\alpha}\vert$ with
$$
<x_{\alpha}\vert x_{\beta }> =\delta_{\alpha\beta}.
$$
In terms of the $N$-fold bracket, we then define
$$
\{ y_0 y_1\dots y_N\} = <y_0\vert \lbrack y_1,y_2,\dots ,y_N\rbrack > .
$$
\par
Zwiebach streamlines the machinery in \cite {KKS}, giving the
classical action in closed string
field theory, the gauge transformations
and showing the invariance of the action.
The classical string action is simply given by
$$S(\Psi ) = {1\over 2} \langle \Psi , Q\Psi \rangle
+ \sum_{n=3}^\infty {\kappa^{n-2} \over n!}
\{ \Psi\dots\Psi \}.
$$
The expression $\{ \Psi\dots\Psi \}$ contains n-terms and will be abbreviated
$\{ \Psi^n \},$ and similarly for $[\Psi^n, \Lambda ]$ below.
\par
The field equations follow from the classical action by simple variation:
$$\delta S = \sum_{n=2}^\infty {\kappa^{n-2} \over n!} n
\{ \delta\Psi , \Psi^{n-1} \}.
$$
The gauge transformations of the theory are given by
$$\delta_\Lambda \vert \Psi >
= \sum_{n=0}^\infty { \kappa^n \over n!}
 [\Psi^n, \Lambda ] .$$
Notice that ALL the terms of higher order are necessary for these to be
consistent.
\par
Similarly, in the sha context (cf. open string field theory as remarked below),
one can define $N+1$-point functions
using the structure maps $m_N$.  A particular example of such a structure
which has quite recently appeared in the physics literature \cite {W}
can be expressed in terms of Massey products:
Let $M$ be a compact oriented
manifold of dimension $m$ and $H$ its deRham cohomology.
Poincar\'e duality gives a graded inner product on $H$ and a pairing
$<\quad\vert \quad >$
between homology and cohomology.  Massey products are determined
by $N$-linear maps $H\otimes \dots\otimes H \to H$ and $N+1$-point
functions then follow.
Because the $N$-linear map has degree $2-N$ and the inner product is
non-zero only for classes whose degrees sum to $m$, the only case in
which all $N$-point
functions $<y y \dots y>$ have the same degree
{\it (and hence can appear simultaneously in a single Lagrangian)}
occurs when the degree of $y$
is 1 and $m=2$.  (A 3-fold Massey product in the complement of the Borromean
rings detects the non-triviality of that link.) {\it With the grading
conventions of closed string field theory, as in} \cite Z, {\it  the
non-polynomial Lagrangian has all fields $\Psi$ of ghost degree 2.}
\subhead Fields: forms versus cohomology classes \endsubhead\par
Recently the point of view of ``cohomological physics'' has become fairly
common.  Theories, both Lagrangian and Hamiltonian, are described initially
in terms of a differential graded vector space or module ( the differential
is often referred to as a {\it BRST operator}), but the physical states
are often cohomology classes, represented by closed ``forms''.  The structures
we have been discussing, whether shLie or shassociative, may occur at either
level \cite R, \cite {HPT}, \cite{S3,5}.
For example, closed string field theory \cite {SZ}, \cite{KKS},
\cite Z
exhibits the structure of an shLie algebra initially at the form level, but
there is an induced structure on cohomology (where
$d_1 =0$, of course).
With $d_1 =0$, the bracket $[\ ,\ ]$ satisfies the Jacobi identity strictly,
but this does NOT imply that the higher order brackets must be zero.
In \cite{WZ}, Witten and Zwiebach describe the cohomology structure only,
but there is one implicit at the form level as well, which Zwiebach works
out very carefully in \cite Z.
\par
In the open string field theory of \cite{HIKKO}, the structure is {\bf homotopy
associative} (the associating homotopy $m_3$ corresponds to the trilinear
operation $(\ \circ\ \circ\ )$ on 3 open string fields), but the higher order
homotopies are zero.    That is, HIKKO define a
convolution  product of string fields  $\Phi * \Psi$  and establish
the relation:
$$
  (\Phi *\Psi ) * \Lambda ) -   \Phi * ( \Psi * \Lambda ) =
 \pm Q(\Phi \circ \Psi\circ \Lambda) \pm (Q\Phi \circ \Psi \circ  \Lambda )
\pm (\Phi \circ Q\Psi \circ \Lambda )
\pm (\Phi \circ \Psi \circ Q\Lambda ),
$$
but, proceeding to four fields, they find (with appropriate signs):

$$
  \Phi *(\Psi \circ \Lambda \circ \Sigma ) \pm (\Phi \circ \Psi \circ \Lambda
)*
\Sigma \pm (\Phi * \Psi)\circ \Lambda \circ \Sigma  \pm \Phi \circ
(\Psi * \Lambda ) \circ \Sigma \pm \Phi \circ \Psi \circ (\Lambda * \Sigma  )
$$
vanishes, rather than being equal to a BRST exact term, cf. (7).
However, there are likely to be non-trivial higher order
N-linear operations on the cohomology.
\par
A reasonably general result that derives higher order operations on
cohomology even when the underlying differential graded algebra is strictly
associative or strictly Lie is given by
Homological Perturbation Theory \cite{HPT}:
\proclaim{Theorem}  Let $(V,d_1)$ be a differential graded vector space of the
same homotopy type as a differential graded algebra (either associative
or Lie) $(A,d_A)$.  A specific homotopy equivalence induces the structure
of an sh (respecitively associative or Lie) algebra on $V$.
\endproclaim

\enddocument

\end